IAC-19-B4.7.11

# Towards End-To-End Design of Spacecraft Swarms for Small-Body Reconnaissance

Ravi teja Nallapu[a], Jekan Thangavelautham*[b]

[a] *Space and Terrestrial Robotics Exploration (SpaceTREx) Laboratory, Department of Aerospace and Mechanical Engineering, University of Arizona, 1130 N Mountain Avenue, Tucson, Arizona, 85721*, rnallapu@email.arizona.edu
[b] *Space and Terrestrial Robotics Exploration (SpaceTREx) Laboratory, Department of Aerospace and Mechanical Engineering, University of Arizona, 1130 N Mountain Avenue, Tucson, Arizona, 85721*, jekan@email.arizona.edu
\* Corresponding Author

**Abstract**

The exploration of small bodies in the Solar System is a high priority planetary science. Asteroids, comets, and planetary moons yield important information about the evolution of the Solar System. Additionally, they could provide resources for a future space economy. While much research has gone into exploring asteroids and comets, dedicated spacecraft missions to planetary moons are few and far between. There are three fundamental challenges of a spacecraft mission to the planetary moons: The first challenge is that the spheres of influence of most moons (except that of Earth) are small and, in many cases, virtually absent. The second is that many moons are tidally locked to their planets, which means that an observer on the planet will have an entire hemisphere, which is always inaccessible. The third challenge is that at a given time about half of the region will be in the Sun's shadow. Therefore, a single spacecraft mission to observe the planetary moon cannot provide complete coverage. Such a complex task can be solved using a swarm approach, where the mapping task is delegated to multiple low-cost spacecraft. Clearly, the design of a swarm mission for such a dynamic environment is challenging. For this reason, we have proposed the Integrated Design Engineering & Automation of Swarms (IDEAS) software to perform automated end-to-end design of swarm missions. Specifically, it will use a sub-module known as the Automated Swarm Designer module to find optimal swarm configurations suited for a given mission. In our previous work, we have developed the Automated Swarm Design module to find swarm configurations for asteroid mapping operations. In this work, we will evaluate the capability of the Automated Swarm module to design missions to planetary moons. We will explore the design space of resonant co-orbits where the spacecraft will have planned periodic encounters with the planetary moon due to the natural dynamics. However, the orientation of the mapping orbits will be a crucial design parameter. Since the arrival trajectories at the target planet do not support captures into any desired inclination, the mission designer should orient the science orbits using the obtained inclination. In this paper, we present a new algorithm to determine the final orientation of the resonant orbits. The proposed algorithm will use a sequence of principal angle rotations to place the apoapsis of the spacecraft's orbit at a desired latitude, longitude, and altitude above the planetary moon. Furthermore, we show that for polar orbits, the algorithm results in a compact solution that can easily determine the orientation of the target orbit. Finally, we will demonstrate the application of the developed algorithm through numerical simulations of a spacecraft swarm mission to map the surface of the Martian moon Deimos.

**Keywords:** Spacecraft swarms, Small body exploration, Resonant co-orbits, Reconnaissance operations, end-to-end mission design

## 1. Introduction

Exploration of small bodies shed fundamental insight into such topics such as the origin of the solar system, the origin of Earth and the origin of life [1, 2]. These bodies are characterized by their small size, irregular shapes, and their corresponding irregular gravity environments. While remote sensing observations of these target bodies from the ground provide useful information, results are limited by the low albedo, low resolution, and atmospheric effects. These factors require missions to get a closer look at these bodies through flybys, orbital insertions, and touch and go missions. The importance of surface exploration of these bodies is also highlighted by the Planetary Science Decadal survey 2013-2022 [3, 4]. Additionally, in-situ missions to Near-Earth Asteroids (NEAs) are being developed to facilitate deep space travel [5]. However, the design of in-situ missions faces some key challenges. First, the physical characteristics of these bodies are poorly understood. Second, the spacecraft dynamics around small bodies constrain the orbits and consumes significant fuel [6]. Therefore, performing detailed reconnaissance without getting into orbit around these bodies is preferred, as they can also allow for touring one or more small bodies. Typically, flyby observations are carried by a single spacecraft equipped with the reconnaissance payload. However, returns from a single observer are limited by both the field of view of the spacecraft instrument, and the duration of the flyby. Additionally, a single monolithic spacecraft may be susceptible to single-point failures. These challenges can be efficiently handled by the swarm approach [7]. An illustration of a swarm operating around a small body is presented in Fig. 1. Design of a swarm mission is a multi-disciplinary problem as it involves the selection of several parameters such as the number of spacecraft, choice of science payload and the design of attitude determination and control, power system, thermal, communications, and propulsion. To simplify the end-to-end design of such a mission requires an integrated design tool and approach. To address these challenges, we developed the Integrated Design Engineering & Automation of Swarms (IDEAS), a software tool to design spacecraft swarms [8]. The IDEAS framework is developed to use evolutionary algorithms that can generate designs that are unintuitive to human designers. An illustration of design using genetic algorithm optimization is shown in Fig. 2. Using genetic algorithms, a form of trial and error learning it is possible to discover novel solutions never thought of by the experimenter and even human competitive solutions utilizing multiple agents [25, 26]. In the IDEAS framework, a swarm mission design is handled by three automated design modules (Fig. 3). This includes an Automated Trajectory Design module, Automated Swarm Design module, and Automated Spacecraft Design module. Each of these modules tries to optimize their respective objectives, and the collective validity of the design is checked by the Mission Analyzer module. The inputs and constraints to the design modules can be provided by a mission designer through the input user interface. This work explores the capability of the Automated Swarm Design module of IDEAS to design swarm flyby missions that generate detailed surface maps of planetary moons. To explore these bodies, the swarm uses resonant co-orbits with the central planet. Here we develop a new algorithm to derive the orientation elements of the resonant to rendezvous with the planetary moon at the desired location. Following this, the swarm design is formulated as an optimization problem which is solved using genetic algorithms. Finally, the Automated Swarm Designer module is demonstrated via numerical simulation of the reconnaissance of the Martian moon Deimos.

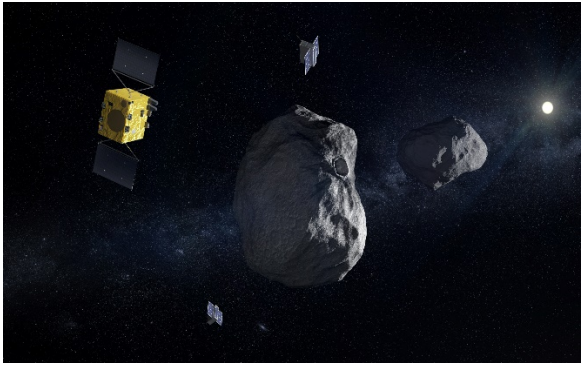

Fig. 1. Illustration of a spacecraft performing reconnaissance around a small body (ESA).

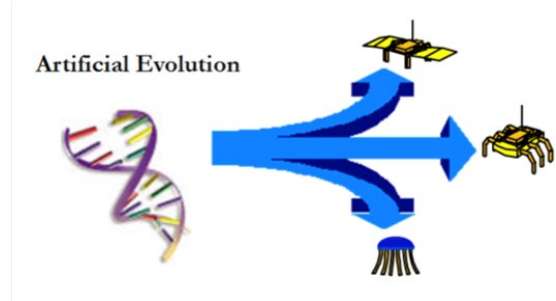

Fig. 2. Illustration of design using evolutionary algorithms. (Image source: JPL)

The organization of the current work is as follows: Section 2 presents related work done in the field of spacecraft swarms. Section 3 presents the methodology. Here we present an overview of the moon mapping mission design problem. In particular, we are interested in swarms that deploy in resonant co-orbits around the central planet to explore the moons. Next, we provide a new algorithm to compute the orientation of co-orbits around the central planet. Using the new algorithm, we compute the orientation of polar resonant co-orbits which the spacecraft in the swarm will use for their reconnaissance. Following this, we formulate a moon mapping mission as an optimization problem which will be solved using a genetic algorithm. Section 4 will demonstrate the capabilities of the algorithms presented to design a surface mapping mission to the Martian moon Deimos. Section 5 will identify and highlight the important contributions of the current work. Finally, Section 6 will conclude the current work by providing conclusions and identifying pathways forward.

## 2. Related Work

Modern work on spacecraft swarms identifies two types of spacecraft architectures: formation flying and constellations [9]. Formation flying spacecraft couple their dynamics in order to maintain specified formations [10], while constellations, on the other hand, require no coordination among their participants [11]. These architectures are also being studied for interplanetary applications [14, 15]. Traditionally, missions are designed based on decoupled design architectures each component of the mission is designed separately [16]. The challenge with these approaches is that individually optimal designs might not be holistic and would often require several iterations to converge to a feasible optimal design. Therefore, a unifying mission design architecture would lead to holistically optimal designs that can provide a better quality of spacecraft missions.

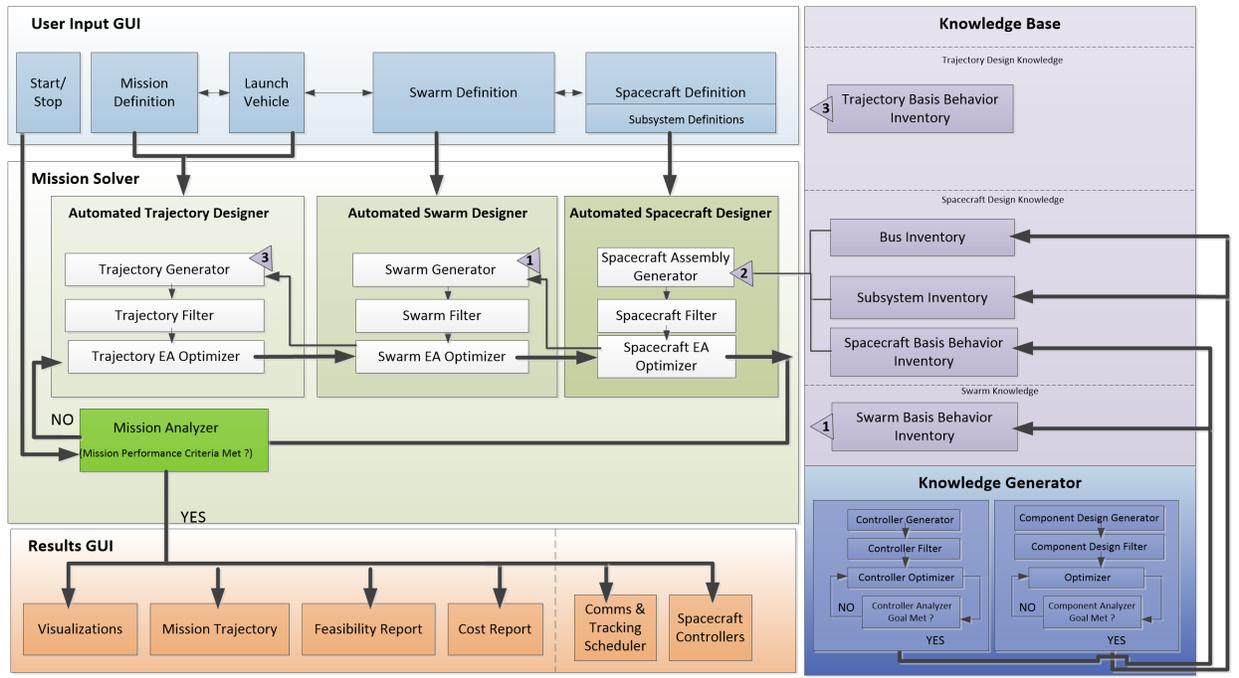

Fig. 3. Software architecture of the proposed IDEAS software to provide an end-to-end design framework for spacecraft swarm missions.

To address these challenges, we developed IDEAS as an end-to-end mission design architecture to design interplanetary spacecraft swarm missions [8]. We then introduced a new classification of spacecraft swarms [17] to bridge the gap between constellations and formation flying swarms. Such a classification allows us to define a unifying scheme for defining swarm architectures. We classified swarms into five classes as follows:

  *Class 0 Swarms*. A Class 0 swarm is a collection of multiple spacecraft that exhibit no coordination either in movement, sensing, or communication.
*Class 1 Swarms*. In a Class 1 swarm, the spacecraft coordinate their movement resulting in formation flying but there is no explicit communication coordination or sensing coordination.
*Class 2 Swarms*. In a Class 2 swarm, the spacecraft coordinate movement and some amount of communication through MIMO or parallel channels. Has sensing but is not optimized to swarm or is post-processed.

*Class 3 Swarms*. A Class 3 swarm coordinates sensing/perception with communication and positioning/movement but doesn't fully exploit the three concurrently. Individual losses can have uneven outcomes include total loss of the system.
*Class 4 Swarms*. Finally, a Class 4 swarm exploits concurrent coordination of positioning/movement, communication and sensing to perform system-level optimization. The system acts if it's a single entity, computing between entity is distributed. Individual losses result in a gradual loss in system performance.

Using these architectures, we demonstrated the capability of IDEAS to design surface mapping swarm missions to asteroids using Class 1 and Class 2 swarm architectures during flybys. We also applied the IDEAS framework to design a Class 0 swarm for monitoring meteor events [18].

The current work will focus on designing Class 2 swarms for exploring planetary moons through the Automated Swarm Designer module of the IDEAS architecture.

2. Methodology

This section presents a requirement-based design of spacecraft swarms to explore planetary moons. We begin by describing the general mission requirements followed by the operations of the Class 2 swarm. The spacecraft will be deployed on polar resonant co-orbits around the central planet. The orbital elements for these co-orbits which will rendezvous with the moon at a specified true anomaly will be derived. Finally, the swarm design will be introduced as an optimization problem.

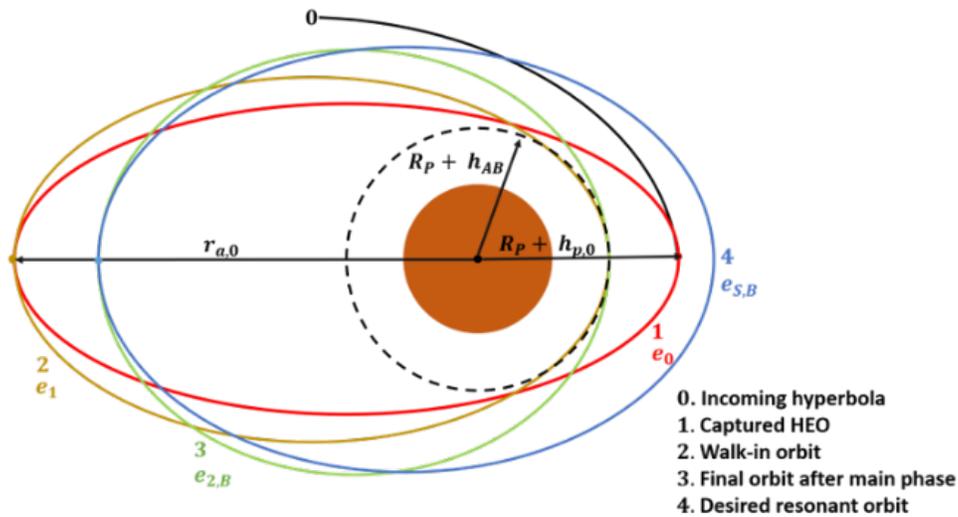

Fig. 4. Anatomy of the aerobraking maneuver showing different phases leading to the final capture around the central planet.

*3.1 Mission definition*

A common mission problem statement for a moon mapping mission would be to generate global surface maps of the moon with a tolerance of $\varepsilon_{map}$, whose observations have a maximum ground resolution of $x_D$, while having a minimum elevation angle of $\varepsilon_D$.

If we assume that the moon has an average radius $R_T$ and that all spacecraft have a camera with an aperture diameter $D_c$, the minimum flyby altitude $h_f$, and spacecraft field of view (FoV) $\eta_D$ to make such an observation can be computed as [16]

$$h_f = \frac{x_D D_C}{\lambda_D} \quad (1)$$

and

$$\sin \eta_D = \left(\frac{R_T}{R_T + h_f}\right) \cos \varepsilon_D \quad (2)$$

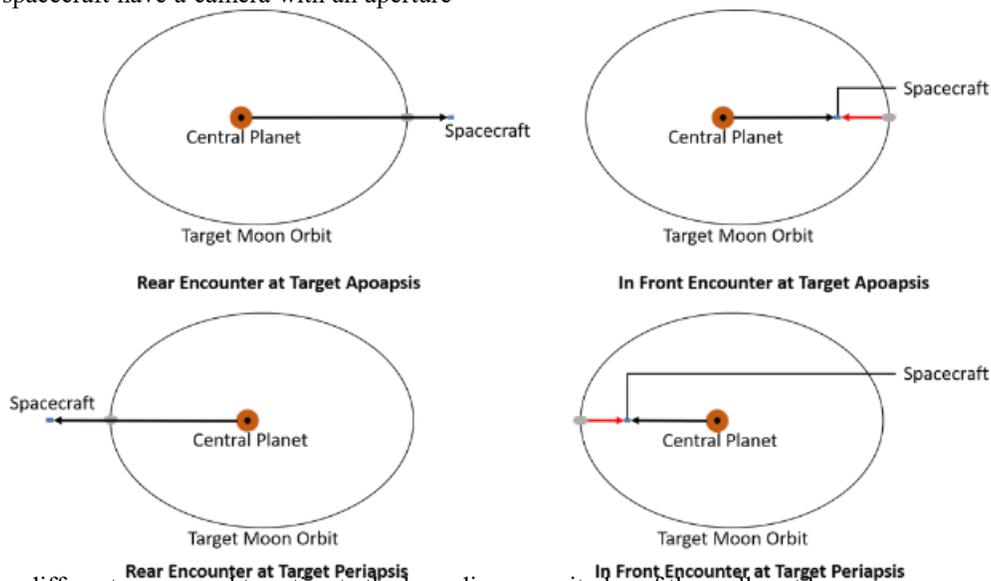

Fig. 5. Four different cases used to estimate the bounding magnitudes of the walk-out burn.

| Specifier | Launch Date | Arrival Date | Periapsis altitude at arrival | Eccentricity of HEO | Resonance | |
|---|---|---|---|---|---|---|
| | | | | | Target body Orbits | Spacecraft Orbits |
| Tr | $D_L$ | $D_A$ | $h_{p,0}$ | $e_0$ | $p$ | $q$ |
| | $\mathbb{R}$ | $\mathbb{R}$ | $\mathbb{R}$ | $\mathbb{R}$ | $\mathbb{Z}$ | $\mathbb{Z}$ |
| | $[D_{L,min}, D_{L,max}]$ | $[D_{A,min}, D_{A,max}]$ | $[h_{p,0,min}, \infty)$ | $[0, e_{0,max}]$ | $[p_{min}, p_{max}]$ | $[q_{min}, q_{max}]$ |

Fig. 6. Gene map of the trajectory design problem showing the design variables involved and their bounds.

The swarm will have to be deployed on co-orbits around the central planet which have a minimum close flyby altitude $h'_f = h_f - \Delta h$ with respect to the moon with $\Delta h > 0$. In this work, we assume that the close encounter flybys with altitude $h'_f$ will occur on the apoapsis of the spacecraft orbit around the central planet.

*3.2 Trajectory design*

The interplanetary trajectory design problem seeks to find trajectories that meet mission constraints such as launch energy $C_3$, the excess velocity at arrival planet $V_{\infty,2}$, and time of flight $ToF$. Traditionally, this done by solving Lambert's problem [19] over a grid of launch dates $D_L$, and arrival dates $D_A$ and then looking at the dispersion of parametric contours on the, well known, porkchop plots [20]. In the current work, we use an optimal search in comparison to the traditional grid search to obtain the optimal trajectory. The objective here will be to minimize the fuel cost $\Delta V$ associated with capturing into an orbit around the central planet. We assume that the central planet has an atmosphere, which will be used to assist the orbit capture.

*3.2.1 Capture with aerobraking*

The aerobraking maneuver is commonly used to reduce the $\Delta V$ associated with orbit capture in an interplanetary mission [19]. This maneuver allows the spacecraft to perform a burn that would barely capture it into a highly eccentric orbit (HEO). After the capture, the spacecraft reduces its eccentricity to its final target value by passing through the planet's atmosphere for virtually no fuel cost. The aerobraking maneuver can be split into four phases as shown in Fig. 4. As shown here, an incoming spacecraft on a hyperbolic trajectory will be captured into an HEO with eccentricity $e_0$ by performing a tangential capture burn of magnitude $\Delta V_C$ at its periapsis located at an altitude of $h_{p,0}$. The spacecraft moves to its apoapsis, where it performs a walk-in burn of magnitude $\Delta V_{WI}$. The walk-in burn moves the spacecraft's periapsis to an altitude $h_{AB}$ above the planet's surface where its atmosphere is strong enough to decelerate the spacecraft. The subsequent phase is called the main phase where the spacecraft reduces its apoapsis altitude by passing through the atmosphere on its periapsis. We assume here that the main phase of aerobraking does not require any maneuvers, while station-keeping maneuvers can be placed to ensure precision. The main phase ends when the apoapsis of the spacecraft orbit reaches that of the target value. Then the spacecraft performs a final walk-out burn at the apoapsis to raise its periapsis to that of the target orbit. Let the magnitude of the walk-out burn be $\Delta V_{WO}$. The total cost of the capture is the sum of the HEO capture, walk-in, and walk-out burns. While the exact value of walk-out burn depends on the final target orbit, we can use four cases to compute its bounding values as shown in Fig. 5. The maximum total cost of the orbit insertion with the aero-assist is therefore expressed as

$$\Delta V_{OI,max} = \Delta V_C + \Delta V_{WI} + \max(\Delta V_{WO}) \quad (3)$$

Where the individual magnitudes of the maneuver costs can be estimated using the Vis-viva energy equation [19]. The values of $\Delta V_{WO}$ are computed for the four cases shown in Fig. 5.

*3.2.2 Resonant orbit shape*

Resonant orbits, in the current context, are orbits with periods that can be expressed as multiples of integer ratios of the moon's period. These orbits enable flyby encounters with a repeating pattern. The semi-major axis of the spacecraft $a_{sc}$ can be computed directly by specifying the two positive integers $p, q$ through Kepler's third law as:

$$\frac{P_T}{P_{sc}} = \frac{p}{q} = \left(\frac{a_T}{a_{sc}}\right)^{\frac{3}{2}} \quad (4)$$

Where $P$ denotes orbital period, $a$ denotes semi-major axis, and the subscripts $T$ and $sc$ denote the correspondence to target moon and the spacecraft respectively. The apoapsis altitude required for imaging the moon at an altitude $h'_f$, and the resonance indices $p, q$ together specify the shape of the resonant orbit.

*3.2.3 Automated trajectory design*
The trajectory design can be posed as a minimization search for trajectories which have the least insertion $\Delta V_{OI,max}$. This can be expressed as

$$\min \Delta V_{OI,max} \quad (5)$$

such that

$$C_3 \leq C_{3,max}$$
$$V_{\infty,2} \leq V_{\infty,2,max}$$
$$ToF \leq ToF_{max}$$

The design variables of the trajectory design problem are presented in Fig. 6 in a gene map format. In addition to the constraints in Equation 5, we place constraints that all peri-apsis altitudes lower bounded by $h_{p,min}$. The bounds for all parameters shown in Equation 5, and Fig. 6 are supplied by a mission designer through the user interface.

*3.3 Swarm design*
The swarm design problem tries to search for the optimal co-orbits which can satisfy the mission requirements described above with a minimum number of spacecrafts. The trajectory design problem from Equation 5 provides the semi-major axis of the resonant orbits. The other fixed orbital elements vary based on the geometry of rendezvous with the moon. While the shape element computation is straight forward, computing the orientation parameters of rendezvous co-orbits is non-trivial. In this section, we first explore the design space of the swarms to map the moon. We then derive the orientation parameters for rendezvous co-orbits and later consider the case for polar co-orbits. Once the shape and orientation parameters are known, the camera coverage of the spacecraft will be computed whenever their distance to the moon falls below the flyby altitude in Equation 1 using camera culling and clipping operations [8]. The attitude of the spacecraft referred to as its behaviour, is very important to compute coverage. In this work, we assume the spacecraft follow the same Class 2 behaviour as described in Reference [17]. In this case, the spacecraft coordinate for communication and orient their communications subsystem towards their leader when they are far from the moon. If their altitude from the moon falls below $h_f$, the spacecraft orient their cameras along their respective line of sight (LoS) to the moon.

*3.3.1 Swarm configuration*
Due to eclipsing constraints from the Sun, the swarm should use multiple visits to get global coverage of the moon. We consider that there are a total of $N_v$ visits of the swarm to the moon. Each of these visits can contain $N_j$ spacecraft and occur when the moon is at a true anomaly of $f_{v,j}$ on its orbit, where $j = 1,2,...,N_v$. An example swarm that has three visits to the moon with a total of nine spacecraft is illustrated in Fig. 7. If the number of visits and number of spacecraft in each visit are known, the swarm size can be computed as

$$N_{sw} = \sum_{j=1}^{N_v} N_j \quad (6)$$

Furthermore, during a visit, the spacecraft should be located differently with respect to the moon to allow coverage variation. For this reason, the close approach of a spacecraft can be specified by its right ascension $\theta_{x,i}$ and declination $\theta_{y,i}$ with respect to the moon as shown in Fig. 8. Here $i = 1,2,...,N_{sw}$.

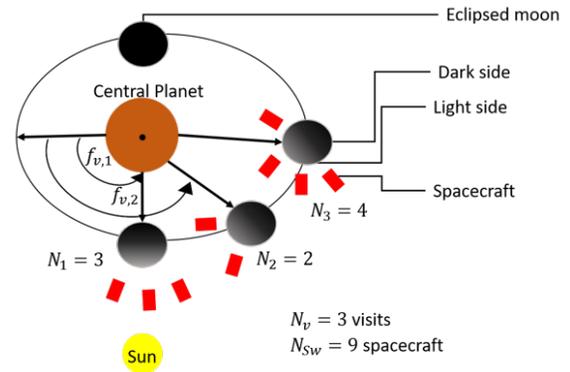

Fig. 7. Example design of a moon mapping swarm.

Specifying, the true anomaly of the moon, and the right ascension and declination angles of the spacecraft during a visit allows us to compute the apoapsis vector of the spacecraft with respect to the central planet, thus allowing us to specify the shape.

*3.3.1 Co orbit orientation*
The right ascension of the ascending node (RAAN), inclination, and argument of periapsis define the 3-1-3 Euler angle rotation set required to describe the orientation of the orbit. Here we present the derivation of the orientation parameters when the true anomaly of rendezvous $f_v$, right ascension $\theta_x$, and declination $\theta_y$ of the visiting spacecraft are specified. Let $\Omega_T$, $in_T$, and $\omega_{p,T}$ represent the RAAN, inclination, and argument of periapsis of the target moon, while $\Omega_{sc}$, $in_{sc}$, and $\omega_{p,sc}$ represent the same for the spacecraft orbits. Additionally, let $R_1$, $R_2$, and $R_3$ be principal rotation matrices about $x, y$, and $z$ axes respectively. The rotation matrix that transforms an inertial planetary frame to the orbit frame of the moon [TP] is given by

$$[TP] = R_3(\omega_{p,T})R_1(in_T)R_3(\Omega_T) \quad (7)$$

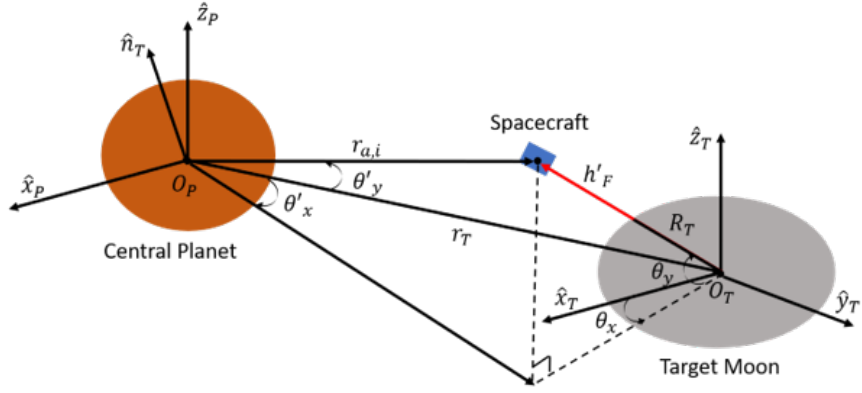

Fig. 8. The geometry of the close encounter of a spacecraft with the target moon.

Let $\bar{r}_a$ be the apoapsis vector of the spacecraft in the planet's inertial frame, constructed from the vector addition of the moon's location at $f_v$, and spacecraft's location of $\theta_x, \theta_y$, and $(R_T + h'_f)$ with respect to the moon. The apoapsis vector in a reference frame rotating with respect to the moon ${}^v\bar{r}_a$ is given by

$$^v\bar{r}_a = R_3(f_v)\bar{r}_a \quad (8)$$

The rotating apoapsis vector is used to extract the planetocentric right ascension $\theta'_x$ and declination $\theta'_y$ angles as

$$\theta'_x = \tan^{-1}\left(\frac{{}^v\bar{r}_a(2)}{{}^v\bar{r}_a(1)}\right) \quad (9)$$

and

$$\theta'_y = \sin^{-1}\left(\frac{{}^v\bar{r}_a(3)}{|{}^v\bar{r}_a|}\right) \quad (10)$$

We can now construct the rotation matrix to transform the planet's inertial frame to the orbital frame of the spacecraft $[SP]$ as

$$[SP] = R_1(-\theta_r)R_2(\theta'_y)R_3(\pi + \theta'_x + f_v)[TP] \quad (11)$$

Where, $\theta_r$ is an unknown auxiliary rotation angle to account for the inclination of the spacecraft orbit. The orientation parameters can be extracted from the rotation matrix [21] in Equation 11 as

$$\tan \Omega_{sc} = \left(\frac{[SP](3,1)}{-[SP](3,2)}\right) \quad (12)$$

$$\tan \omega_{p,sc} = \left(\frac{[SP](1,2)}{[SP](2,3)}\right) \quad (13)$$

and

$$\cos in_{sc} = [SP](3,3) \quad (14)$$

It is now evident that Equations 12-14 depend on the auxiliary rotation angle $\theta_r$. Evaluating Equation 14 by examining the elements of the matrix $[SP]$ gives us

$$\cos in_{sc} = \cos \theta_r \left(\cos in_T \cos \theta'_y - \sin in_T \sin \theta'_y \sin \alpha\right) + \sin \theta_r \left(-\sin in_T \cos \alpha\right) \quad (15)$$

Where,

$$\alpha = \theta'_x + f_v + \omega_{p,T} \quad (16)$$

| Specifier | Number of Visits | Number of spacecraft during each visit | | | True anomaly of the target moon during each visit | | | | RA angles of spacecraft during each visit | | | | Dec angles of spacecraft during each visit | | | |
|---|---|---|---|---|---|---|---|---|---|---|---|---|---|---|---|---|
| | $N_v$ | $N_1$ | $N_2$ | ... | $N_{N_v}$ | $f_{v,1}$ | $f_{v,2}$ | ... | $f_{v,N_v}$ | $\theta_{x,1}$ | $\theta_{x,2}$ | ... | $\theta_{x\,N_{sw}}$ | $\theta_{y,1}$ | $\theta_{y,2}$ | ... | $\theta_{y\,N_{sw}}$ |
| SwM | $\mathbb{Z}^+$ $[1, N_{v,max}]$ | $\mathbb{Z}^+$ $[1, N_{j,max}]$ | | | | $\mathbb{R}$ $[0, 2\pi]$ | | | | $\mathbb{R}$ $[0, 2\pi]$ | | | | $\mathbb{R}$ $[-\frac{\pi}{2}, \frac{\pi}{2}]$ | | | |

Fig. 9. Gene map of the mapping swarm design problem showing the design variables involved and their bounds.

Equation 15 can be solved for the auxiliary rotation angle either analytically or numerically if the inclination of the spacecraft $in_{sc}$ is known.

*3.3.2 Polar co-orbits*
The computations presented in Equation 15 can be simplified if we assume that the swarm will only enter polar orbits around the planet. Substituting $in_{sc} = \frac{\pi}{2}$ in Equation 15 allows us to compute $\theta_r$ as

$$\tan \theta_r = \frac{\cot in_T \cos \theta'_y - \sin \theta'_y \sin \alpha}{\cos \alpha} \quad (17)$$

This allows us to compute the orientation of polar co-orbits if $f_v$, $\theta_x$, and $\theta_y$ are specified.

*3.3.4 Automated swarm design*
The swarm design problem can be posed as an optimally sized swarm that can generate the required coverage. The coverage is computed as a set of culling and clipping operations with respect to the spacecraft-moon LoS [8]. Only the observations of the illuminated side are taken into consideration for computing the percentage of the total surface area observed $P_{obs}$. The illumination is modelled as a culling operation with respect to the Sun-moon LoS vector. Therefore, the automated swarm design problem is posed as

$$\min N_{sw} \quad (18)$$

such that

$$100 - P_{obs} \leq \varepsilon_{map}$$

Additionally, designs that result in resonant orbits that either collide with the moon or with among other spacecraft will be filtered out by an additional constraint. Equations 5 and 18 will be solved using a mixed-integer genetic algorithm optimizer [22]. The gene map showing the design variables of the swarm and their bounds is shown in Fig. 9.

**4. Numerical simulations and results**

This section demonstrates the algorithms described above through numerical simulations by designing a mission to explore the Martian moon Deimos. The example mission objective that we consider is as follows.

*4.1 Mission objective*
We are interested in designing a minimum sized swarm that can map at least 99 % of Deimos. The images have a maximum ground resolution of $1\ m/px$ and a minimum elevation angle of 30 deg.

A 5000 polygon model of Deimos [23] was used to model the shape of Deimos. The orbit of Deimos [24] and the orbit of Mars around the Sun [19] are programmed into the IDEAS framework.

*4.2 Spacecraft design*
A database of small spacecraft parts is programmed into the IDEAS framework. The database is used to generate the spacecraft subsystems and is also used to estimate the cost using appropriate spacecraft cost models [16]. In the current work, we assume that the spacecraft in the swarm are $27U$ CubeSats. The payload of the spacecraft is an 8 cm aperture visible camera.

*4.3 Trajectory design*
As described above, the trajectory of the swarm from Earth to Mars is obtained by solving Equation 5 through a genetic algorithm optimizer. The following parameters used as inputs to the solver are presented in Table 1.

*4.3.1 Optimal trajectory*
The optimizer searched for nearly 600 generations of solution where each solution spanned 200 solutions. The selected optimal solution was typically located within the first 200 generations of search. The problem was solved multiple times to verify the convergence of the solution. The results of the trajectory design optimizer showing the mean and best designs along with the selected optimal solution are presented in Fig. 10. As seen here, the optimal solution suggests that the swarm should be launched from Earth on 9th Aug 2020 and would arrive at Mars on 25th Aug 2020. This 200 day trip requires a $C_3$ energy of $17.6\ km^2/s^2$ at launch and arrives at Mars with a $V_{\infty,2}$ of $2.49\ km/s$. The optimal resonant orbit for the swarm is an $8{:}18$ resonance, which requires a maximum insertion $\Delta V$ of 0.66 km/s.

*4.4 Swarm design*
The mapping swarm is designed by solving Equation 18 using a mixed-integer genetic algorithm optimizer. The Heliocentric orbit of Mars and Martian orbit of Deimos is propagated for one orbital period using central body gravity models. During this orbit, the coverage of the swarm is noted.

*4.4.1 Optimal swarm*

For the optimal swarm search problem, the algorithm was able to converge to an optimal solution within 20 generations, exploring a total of 2000 designs. The solver was run multiple times to verify the convergence. The results of the swarm design optimization along with the selected optimal solution is presented in Fig. 11. As seen here, the optimal swarm contains a total of 11 spacecraft that would visit Deimos in three separate locations. The true anomaly of Deimos during the visit, right ascension and declination of spacecraft during close encounters during the visit are also presented in Fig. 11.

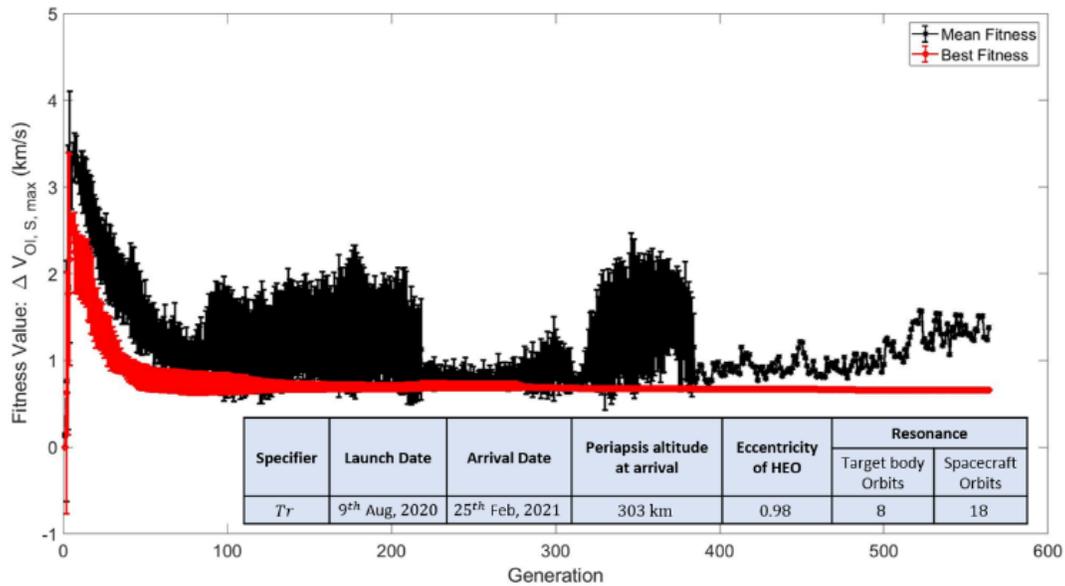

Fig. 10. The output of the trajectory design algorithm showing the evolution of mean and best fitness across multiple generations along with the selected optimal solution.

Table 1. Input parameters to the trajectory design problem.

| Parameter | Value |
| --- | --- |
| Launch date range | 1/1-31/12/2020 |
| Arrival date range | 1/5/2020-31/12/2021 |
| Min periapsis altitude | 300 km |
| Max HEO eccentricity | 0.98 |
| Deimos resonance bounds, $[p_{min}, p_{max}]$ | [1, 10] |
| S/C resonance bounds, $[q_{min}, q_{max}]$ | [1, 20] |
| Max launch energy | 20 km$^2$/s$^2$ |
| Max excess velocity | 3 km/s |
| Max time of flight | 200 days |
| Imaging altitude offset, $\Delta h$ | 5 km |
| Aerobraking altitude, $h_{AB}$ | 150 km |

*4.4.2 Mapping performance*
The resonant co-orbits system of the swarm around Mars is presented in Fig.12. The three rendezvous locations of the swarm can be seen here. Each of these three visits involves the corresponding number of spacecraft flying by Deimos while imaging its illuminated side. The performance of the mapping operation is visualized in Fig. 13. The designed swarm is able to map 99.1 % of the Deimos' surface as presented in Fig. 13.

## 5. Discussion

The current work demonstrated a few practical challenges associated with interplanetary swarm mission design. Firstly, such mission designs are highly multidisciplinary. In many cases, the design of spacecraft constrains the trajectory of the spacecraft, which in turn, influences the performance of the swarm. Next, the design of swarm missions to moons is challenged by dynamic constraints such as illumination, rotation, tidal locking, and irregular shapes. Due to such constraints, a swarm mission design to small bodies can be very unintuitive to human designers. For these reasons a multi-disciplinary optimization tool such as IDEAS, which uses evolutionary algorithms to optimize the designs, would result in holistically optimal missions with better performance.

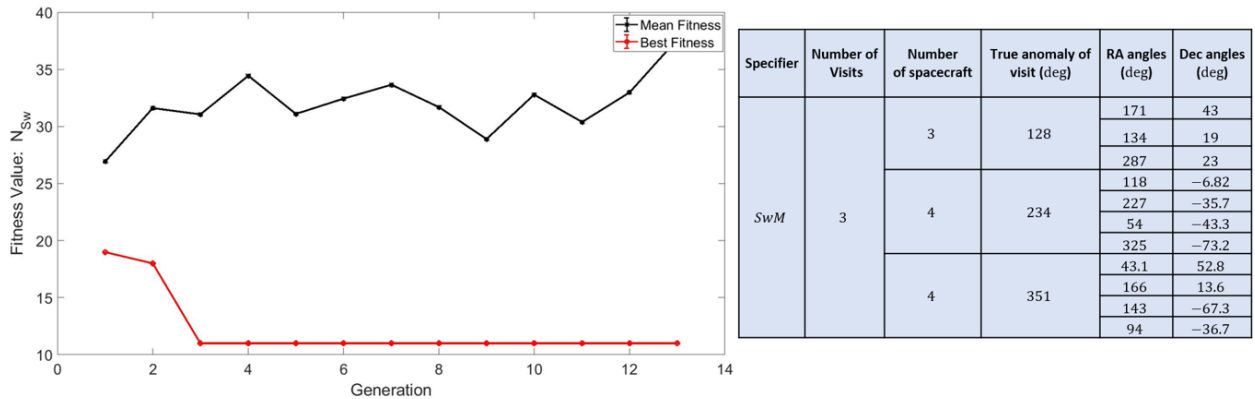

Fig. 11. Result of the mapping swarm optimizer showing the evolution of the mean and best solutions across different generations (left) along with the selected optimal solutions (right).

The following are a few important contributions of the current work to the state-of-the-art mission design. The first was to develop a new architecture to design swarm missions to planetary moons. Next, we developed a new algorithm to determine co-orbits that rendezvous with a planetary moon at a specified location. Finally, we demonstrated how the swarm mission design is a multidisciplinary problem and developed a new architecture to design end-to-end missions of spacecraft swarms.

## 6. Conclusions

The current work explored the development of an end-to-end architecture to explore small bodies in the solar system using spacecraft swarms. While our previous work explored the mission design cases to asteroids, the current work discussed the capabilities of IDEAS to explore planetary moons. The swarm, in the current work, would enter into resonant co-orbits around the central planet to get periodic visits to the moon. Here we developed a new algorithm to determine general co-orbits that lead to rendezvous with the moon. We then derived a closed-form solution for the orientation of these orbits if they were polar. We then formulated the mission design as a set of multi-disciplinary optimization problems. Finally, we demonstrated all the algorithms and architectures developed in the current work through numerical simulations.

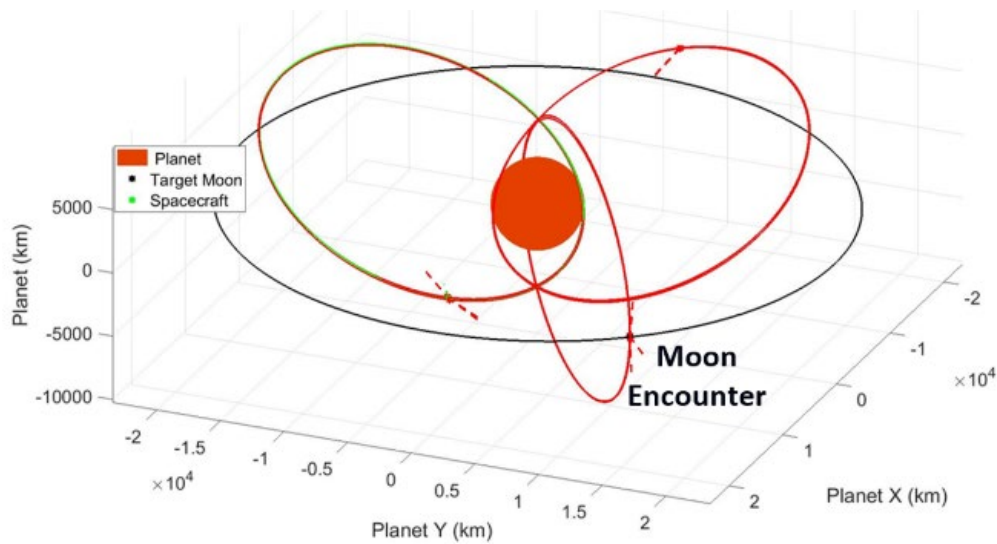

Fig. 12. Optimal resonant co-orbits of the swarm around Mars to map Deimos during a swarm visit.

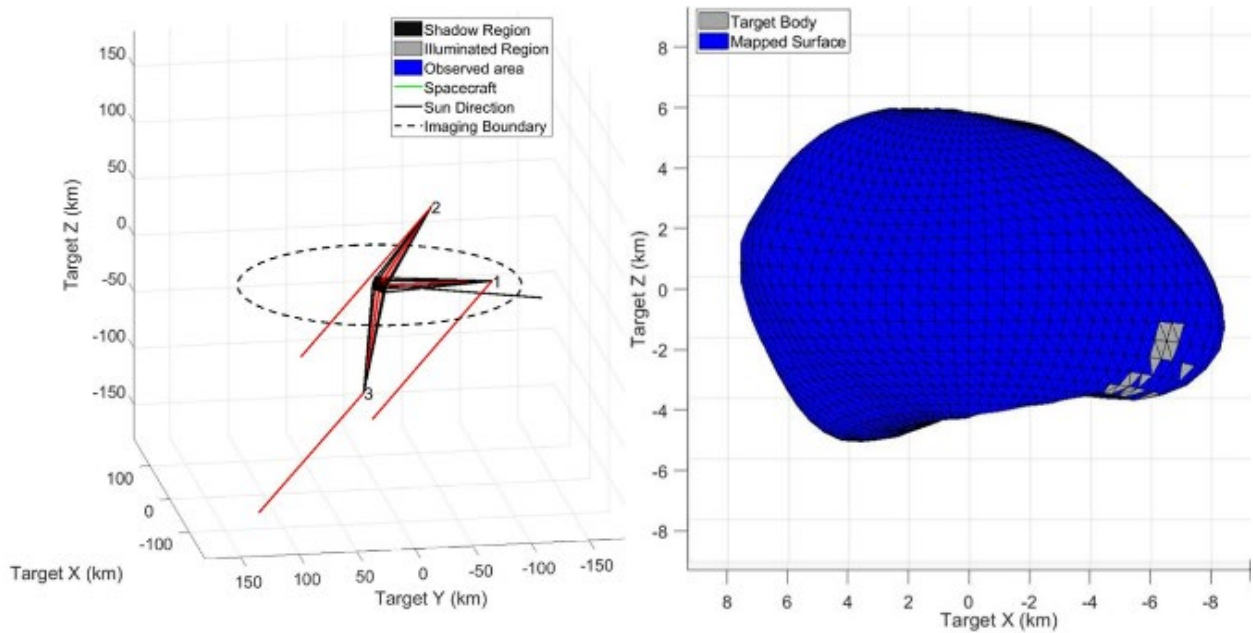

Fig. 13. Results of the mapping mission showing one of the three close encounter flybys of the swarm (left), and the surface that was mapped by the swarm after all three passes (right).

Our future work on IDEAS will focus on embedding high fidelity dynamical models into the simulations. Such perturbations can be used to accurately estimate the fuel cost of spacecraft maneuvers and their placements. Additionally, we will also focus on developing the Automated Spacecraft Designer module of the IDEAS framework by maintaining a regularly updated database of spacecraft subsystem components. This will enable IDEAS to optimize spacecraft, trajectory, and the swarm performance simultaneously making it an end-to-end tool for designing swarm missions to solar system small bodies.